\documentclass[prb, twocolumn]{revtex4-2}
\usepackage{bm}
\usepackage{graphicx}
\usepackage[x11names]{xcolor} 

\usepackage{amsmath}
\usepackage{bbold}
\usepackage{mathrsfs} 
\usepackage[unicode=true,colorlinks=true]{hyperref} 
\hypersetup{urlcolor = blue }

\newcommand{\braket}[3]{\left\langle #1 \left| #2 \right| #3 \right\rangle}

\begin{document}

\title{Exciton interaction with acoustic phonons in PbS nanocrystals}
\author{M.O.~Nestoklon$^{1}$}
\author{S.V.~Goupalov$^{1,2}$} 
\email{serguei.goupalov@jsums.edu}
\affiliation{$^1$~Ioffe Institute, 194021 St. Petersburg, Russia\\
$^2$~Department of Physics, Jackson State University, Jackson MS 39217, USA}

\begin{abstract}
The interaction between exciton and acoustic phonons via deformation potential in PbS nanocrystals is calculated in the ${\bf k} \cdot {\bf p}$ model. The size-dependence of the Huang-Rhys factors for the fundamental breathing mode and its overtones as well as for spheroidal vibrations with the total angular momentum $j=2$ are evaluated.
\end{abstract}

\maketitle

\section{Introduction}
 
Lead chalcogenide (PbX, X=S, Se, Te) nanostructures are widely used in optoelectronics~\cite{Garuge08,Sukhovatkin09,Lu2020} due to the fact that their optical properties can be tuned within the near-infrared and mid-infrared ranges. These materials have both conduction and valence band extrema located at the four $L$ points of the Brillouin zone forming four inequivalent anisotropic valleys \cite{Kang1997}. A high degeneracy of the energy spectrum in bulk materials gives rise to a complex fine structure of exciton levels in PbX quantum dots (QDs)~\cite{Avdeev2020} which can be probed by single-QD spectroscopy at cryogenic temperatures~\citep{Hu19}.

Vibrational modes of QDs corresponding to confined acoustic phonons can be observed in low-frequency Raman scattering and in time-domain experiments,
where they can be excited in a coherent way~\cite{krauss_prl,Ikezawa2001}. Although symmetry of QDs is lower than spherical, one can usually describe their low-frequency vibrations in terms of Lamb modes of an elastic sphere~\cite{lamb1882,saviotreview,ufn}. In first experiments on PbS colloidal QDs~\cite{krauss_prl,krauss_prb}, basing on polarization properties of Raman scattering and frequency assignment of the Lamb modes, it was determined that the phonon mode dominating low-frequency Raman spectra and excited in time-domain experiments is the fundamental radial breathing mode. However, later experiments on PbSe QDs embedded in phosphate glass~\cite{Ikezawa2001} have revealed that low-frequency Raman spectra are dominated by the spheroidal vibrational mode corresponding to the total angular momentum $j=2$. 

Macroscopically, selection rules for Raman scattering are determined by the symmetry properties of the second-order susceptibility for the low-frequency excitation~\cite{Loudon}. For a spherical QD it follows that
both the radial breathing mode having $j=0$ and the spheroidal mode with $j=2$ are Raman active~\cite{duval,Goupalov2006}. Therefore, in order to verify experimental claims, one has to consider a particular microscopic mechanism of exciton-phonon coupling. 

The 
strength of the exciton coupling with acoustic phonons in lead chalcogenide QDs has recently attracted attention, since single-dot experiments have revealed that exciton dephasing is responsible for homogeneous broadening of photoluminescence (PL) spectral lines~\citep{Hu19}. Huang-Rhys factors describing this strength for various modes were used as adjustable parameters in Ref.~\citep{Hu19} in order to fit experimentally observed PL lineshapes.  While the strength of the exciton coupling to the radial breathing mode has been estimated in Ref.~\cite{krauss_prl}, no such estimation, to the best of our knowledge, has been attempted for spheroidal phonons with $j=2$. In this paper we will fill this gap and also revisit the question of exciton coupling to the radial breathing mode.

We will describe charge carriers confined in spherical QDs using the theory, developed by Kang and Wise \cite{Kang1997}, where the 
electron and hole states are characterized by the four-component wave functions. In the lowest order of this theory the valley states are isotropic and are described in terms of effective masses and interband momentum matrix elements averaged over the longitudinal and transverse (with respect to the valley axis) values~\cite{Kang1997}.

The rest of the paper is organized as follows. In Sec.~\ref{sec:matrel} we introduce the model and perform a symmetry analysis.
In Sec.~\ref{sec:rbm} we consider the radial breathing mode and calculate the Huang-Rhys factor describing 
the strength of the exciton-phonon interaction
for this mode. In Sec.~\ref{sec:sph1} we consider exciton interaction with the spheroidal phonon mode characterized by the total angular momentum $j=2$.
In Sec.~\ref{concl} we present concluding remarks.

For numerical calculations we will use 
PbS QDs since material parameters for this compound are best known.


    \section{Matrix elements of the electron-phonon interaction}\label{sec:matrel}

In the isotropic model of Kang and Wise~\cite{Kang1997} (also known as the spherical Dimmock model), the wave function of a charge carrier originating from each $L$-valley can be represented as a bispinor satisfying the following equation ($\hbar=1$, $m_0=1$)~\cite{Kang1997}
\begin{equation}
\label{dimmock}
\begin{pmatrix}
\frac{E_g}{2} - \alpha_c \, \Delta 
&
-i \, P \left( {\bm \sigma} {\bm \nabla} \right)
\\
-i \, P \left( {\bm \sigma} {\bm \nabla} \right)
&
- \frac{E_g}{2} + \alpha_v \, \Delta 
\\
\end{pmatrix}
\,
\begin{pmatrix}
\hat{u} \\
\hat{v}
\end{pmatrix}
=E \, 
\begin{pmatrix}
\hat{u} \\
\hat{v}
\end{pmatrix}
\,.
\end{equation}
Here $\hat{u}$ and $\hat{v}$ are spinors, $\sigma_{\beta}$ ($\beta=x,y,z$) are the Pauli matrices, $\alpha_c$, $\alpha_v$, $E_g$, and $P$ are parameters of the model, and $E$ is the electron energy.

The electron and hole states of charge carriers confined in a QD with radius $R$ are found from the condition  that a linear combination of the two linearly independent solutions of Eq.~(\ref{dimmock}), given explicitly in Appendix~\ref{sec:KW} by Eqs.~(\ref{sol1}), (\ref{sol2}), vanishes at the QD surface. 
The lowest electron state in the conduction band with the angular momentum projection $M_c=\pm1/2$ is characterized by the bispinor
\begin{subequations}\label{eq:kp_wf}
\begin{equation}
  \hat{\psi}_{M_c}^{c}({\bf r}) =  \frac{1}{R^{3/2}} \, \begin{pmatrix}
      z_0^c(x) \, \hat{\Omega}^0_{1/2,M_c}(\bm{r}/r) \\ 
    i \, z_1^c(x) \, \hat{\Omega}^1_{1/2,M_c}(\bm{r}/r) \\ 
  \end{pmatrix} \,,
\end{equation}
where $x \equiv r/R$ and $\hat{\Omega}^l_{1/2,M_c}(\bm{r}/r)$ is the spherical spinor~\cite{Varshalovich} ($l=0,1$).
The ground band-edge hole state is characterized by the bispinor
\begin{equation}
 \hat{\psi}_{M_h}^{h}({\bf r}) = \frac{1}{R^{3/2}} \, \begin{pmatrix}
        z_1^v(x) \, \hat{\Omega}^1_{1/2,M_h}(\bm{r}/r) \\ 
      - i \, z_0^v(x) \, \hat{\Omega}^0_{1/2,M_h}(\bm{r}/r) \\ 
  \end{pmatrix} \,.
\end{equation}
\end{subequations}
Explicit form of the functions $z_{l}^{\eta}(x)$ with $l=0,1$ (which implicitly depend on the QD radius $R$), together with the dispersion equations,
allowing one to find the energies of quantum confined states in QDs, are presented in Appendix~\ref{sec:KW}. Using these functions, the matrix elements of the electron-phonon interaction can be calculated as
\begin{multline}\label{eq:eph_defin}
  \braket{\eta,N}{\hat{H}_{def}}{\lambda,M} \\= 
  \int{\mathrm{d}{\bf r} \left(\hat{\psi}_N^{\eta}({\bf r}) \right)^+ 
  \begin{pmatrix} \hat{H}_{def}^c({\bf r}) & 0 \cr 0 & \hat{H}_{def}^v({\bf r}) \end{pmatrix} \hat{\psi}_M^{\lambda}({\bf r})} \,.
\end{multline}
Here $\hat{H}_{def}^{\eta}({\bf r})$ stems from deformations produced by phonon modes. 
The general form of the deformation-induced change of the energy minima (maxima) of the conduction (valence) band in the multivalley cubic semiconductors may be written as \cite{Balslev66,Walle86}:
\begin{equation}\label{eq:Xi_defin}
  \left(\hat{H}_{\mathrm{def}}^{\eta}\right)_{ij} = 
  \sum_{\alpha\beta} \left[ \Xi_d^{\eta} \delta_{\alpha\beta} + \Xi_u^{\eta} n^i_{\alpha}n^j_{\beta} \right] \hat{u}_{\alpha\beta} 
\end{equation}
where $\bm{n}^i$ is the direction vector of the $i$-th valley, the index ${\eta}=c$($v$) refers to the conduction (valence) band, and $\hat{u}_{\alpha \beta}$ is the strain tensor.

In this work we will apply Eq.~\eqref{eq:Xi_defin} to the L-valley in the coordinate frame with the $z$ axis along the valley wave vector ${\mathbf{k}}$. In this case, the interaction Hamiltonian takes the form~\cite{herring,apy,gl}
\begin{equation}\label{eq:D_defin}
\hat{H}_{\mathrm{def}}^{\eta}(\bm{r})=\Xi_d^{\eta} \left( \hat{u}_{xx} + \hat{u}_{yy} + \hat{u}_{zz} \right) + \Xi_u^{\eta} \hat{u}_{zz} \,,
\end{equation}
where $\Xi_d^{\eta}$ and $\Xi_u^{\eta}$ are the dilatation and uniaxial deformation potentials, respectively. 
This Hamiltonian 
is diagonal in spin indices. Strictly speaking, when $\hat{H}_{\mathrm{def}}^{\eta}(\bm{r})$ enters expressions like Eq.~(\ref{eq:eph_defin}), it should be multiplied by a 
$2 \times 2$ unit matrix which we omit for brevity, as we did in Eq.~(\ref{dimmock}).
The Hamiltonian~(\ref{eq:D_defin}) 
can be expanded via spherical harmonics:
\begin{equation}
    \hat{H}_{def}^{\eta}(\bm{r}) = \sum_{\ell,\ell_z} \hat{D}^{\eta}_{\ell \ell_z}(r) Y_{\ell \ell_z}(\bm{r}/r)\,.
\label{expansion}    
\end{equation}

Spheroidal (as opposed to torsional) vibrational modes of a sphere 
are characterized by the total angular momentum $j$ and the parity $(-1)^{j+1}$~\cite{ufn}. Therefore, the deformation potential~(\ref{eq:D_defin}) is an even function of the coordinates for the spheroidal modes with $j=0,2$.
The spherical spinors entering Eq.~(\ref{eq:eph_defin}) are characterized by the total angular momenta $1/2$ and transform according to the corresponding irreducible representations of the full rotational group. Therefore, only terms proportional to $Y_{00}$ in Eq.~(\ref{expansion}) will contribute to the integral of Eq.~(\ref{eq:eph_defin}).
Then it follows that
\begin{equation}
  \braket{\eta,N}{\hat{H}_{def}}{\lambda,M} = \hat{\Delta}^{\eta} \delta_{\eta\lambda}\delta_{MN} \,,
\end{equation}
where
\begin{multline}\label{eq:eph_matrel_res}
  \hat{\Delta}^{\eta}=\frac1{\sqrt{4\pi}}\int_0^1 \left[ 
    \hat{D}_{00}^{\eta}(Rx) \left(z_0^{\eta}(x)\right)^2 + 
    \right.
\\
\left.    
    \hat{D}_{00}^{\bar{\eta}}(Rx) \left(z_1^{\eta}(x)\right)^2
  \right]  x^2 \mathrm{d}x  
   \,.
\end{multline}
Here $\bar{\eta}=c$ when $\eta=v$ and $\bar{\eta}=v$ when $\eta=c$.
We note that the arguments given above confirm the general selection rules for Raman scattering from spherical particles~\cite{duval,Goupalov2006} for the case of this particular model and interaction mechanism.

\section{Huang-Rhys factor for the radial breathing mode}\label{sec:rbm}
The properly normalized displacement corresponding to the radial breathing mode has the form (see Appendix~\ref{sec:ph_br})
\begin{widetext}
\begin{equation}\label{eq:ph_ampl_b}
\hat{{\bf u}}({\bf r})= \omega_0 \, \sqrt{\frac{\hbar}{8 \pi \rho c_l^3}} \,
\frac{j_1(qr) \, {\bf e}_r}{\sqrt{\frac{qR}{2}-\frac{\sin^2{qR}}{qR}+\frac{\sin{qR} \, \cos{qR}}{2}}} \,
\left( \hat{a}^{\dag} +\hat{a} \right) 
\equiv \hat{A}_{0} \, j_1(qr) \, Y_{00}(\bm{r}/r) \, {\bf e}_r \,,
\end{equation}
\end{widetext}
where $\omega_0$ is the frequency of the breathing mode, $\rho$ is the mass density of PbX, $j_1(x)$ is the spherical Bessel function of the $1$st order, 
${\bf e}_r$ is the radial unit vector, the value of $qR \approx 2.95$ is determined from the dispersion equation, see \cite{ufn}.
$\hat{a}^{\dag}$ and $\hat{a}$ are the phonon creation and destruction operators, $\hat{A}_{0}$ is introduced for convenience in order to shorten notations.

The interaction Hamiltonian \eqref{eq:D_defin} in the conduction or valence band takes the form ($z||[111]$):
\begin{multline}\label{eq:HdefBM}
\hat{H}^{\eta}_{def}({\bf{r}}) = D_{\mathrm{av}}^{\eta} \bm{\nabla\cdot \hat{u}} + 
D_{\mathrm{an}}^{\eta}
\left(\frac{\partial{\hat{u}_x}}{\partial x} + \frac{\partial{\hat{u}_y}}{\partial y} -2  \frac{\partial{\hat{u}_z}}{\partial z} \right) 
= \\
D_{\mathrm{av}}^{\eta}  \, \hat{A}_0 \, q \, j_0(qr) \, Y_{00}(\bm{r}/r) 
-\frac2{\sqrt5} D_{\mathrm{an}}^{\eta}\, \hat{A}_0 \, q \, j_2(qr) \, Y_{20}(\bm{r}/r) \,,
\end{multline}
where we introduced the average $D_{\mathrm{av}}^{\eta}=\Xi_d^{\eta}+\frac13\Xi_u^{\eta}$ and the anisotropic $D_{\mathrm{an}}^{\eta}=-\frac13\Xi_u^{\eta}$ parts of the deformation potential. The effective Hamiltonian of the electron-phonon interaction is calculated as the matrix elements of the Hamiltonian \eqref{eq:HdefBM} taking into account that, as was demonstrated in the previous section, only terms proportional to $Y_{00}$ contribute to the result \eqref{eq:eph_matrel_res}.

The Huang-Rhys factor is given by
\begin{equation}\label{eq:HRb}
S_0=\frac{(\Xi^c-\Xi^v)^2}{\hbar^2 \omega_0^2} \,,
\end{equation}
where 
\begin{equation}
  \Xi^{\eta} \, \left( \hat{a}^{\dag} +\hat{a} \right) \equiv \hat{\Delta}^{\eta} \,,
\end{equation}
and $\hat{\Delta}^{\eta}$ is caclulated using Eq.~\eqref{eq:eph_matrel_res} with
\begin{equation}\label{eq:D00_b}
  \hat{D}_{00}^{\eta,\mathrm{b}} = D_{\mathrm{av}}^{\eta}  \, \hat{A}_0 \, q \, j_0(qr)\,.
\end{equation}

It is convenient to represent the Huang-Rhys factor as
\begin{equation}\label{eq:HR_b}
 S_0 = \frac{\mathcal{S}_b }{R^2} \left[ \mathscr{D}(D_{\mathrm{av}}^c,D_{\mathrm{av}}^v,\xi_0) \right]^2\,,
\end{equation}
where 
\begin{equation}
 \mathcal{S}_b = \frac1{8\pi\hbar\rho c_l^3}\frac{\xi_0^2}{\frac{\xi_0}2-\frac{\sin^2\xi_0}{\xi_0}+\frac{\sin\xi_0\cos\xi_0}2  }
\end{equation}
is independent of the QD size
and 
\begin{multline}
 \mathscr{D}(D^c,D^v,\xi) =\int
 \Big[
   D^c\left(z_0^c(x)\right)^2
  +D^v\left(z_1^c(x)\right)^2
\\
  -D^v\left(z_0^v(x)\right)^2
  -D^c\left(z_1^v(x)\right)^2
 \Big]  x^2 j_0(\xi x) \mathrm{d}x
\end{multline}
depends on the QD size only implicitly through the functions $z_l^{\eta}$.
Here $\xi_0=qR$ is found from the transcendental equation~\cite{ufn} $[1-(\xi_0 c_l/2c_t)^2]\sin\xi_0=\xi_0\cos\xi_0$. For PbS material parameters, the numerical values of $\mathcal{S}_b$ at the temperatures 4~K and 300~K for the fundamental breathing mode ($n=0$) and its first overtone ($n=1$) are given in Table~\ref{tbl:prefactors}.

Now let us discuss material parameters.
For the sound velocities we refer to the work of Chudinov~\cite{chudinov} where temperature dependences of longitudinal and transverse sound velocities for waves propagating along [100] in bulk PbS were measured in the temperature range from $77$~K to $650$~K. We extrapolate these linear temperature dependences to the entire temperature range and use the corresponding sound velocities as the parameters of the isotropic elastic model. This gives
\begin{align}
\label{cl}
c_l(T)&=\left(4.408 \cdot 10^5 - 108 \cdot T \right) \mbox{cm/s} \,,
\\
\label{ct}
c_t(T)&=\left(1.607 \cdot 10^5 - 37 \cdot T \right) \mbox{cm/s} \,,
\end{align}
where $T$ is in Kelvins.

With these parameters, for a QD with $R=15$~\AA\mbox{} the frequency $\omega_0=c_l q$ of the radial breathing mode is
$\hbar \omega_0=5.28$~meV at $T=300$~K and $\hbar \omega_0=5.69$~meV at $T=4$~K.

\begin{figure}[thb]
\includegraphics[width=\linewidth]{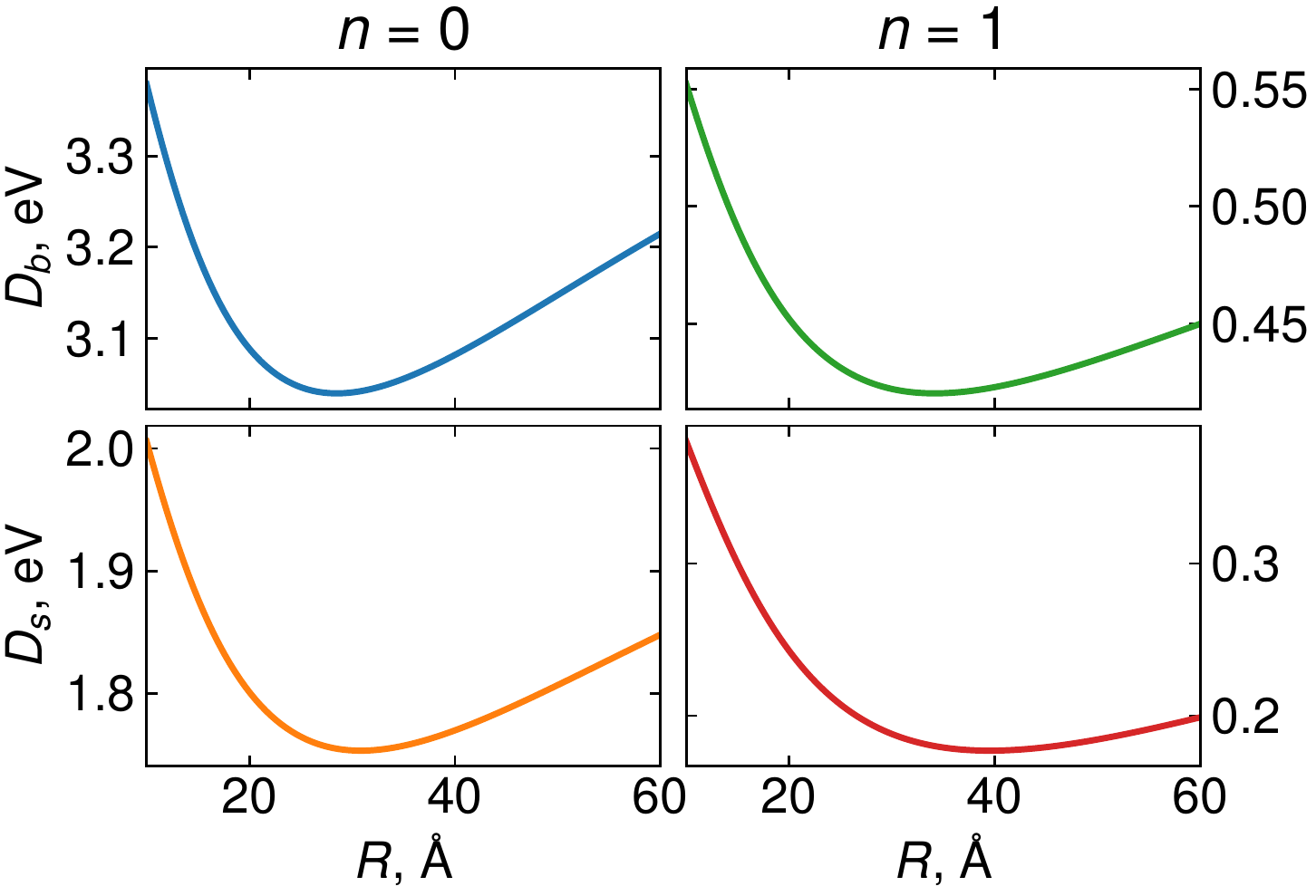}
\caption{Upper panels show the size dependence of the $D_b \equiv \mathscr{D}(D_{\mathrm{av}}^c,D_{\mathrm{av}}^v,\xi_0)$ for fundamental breathing mode (left) and first overtone $n=1$ (right) calculated using $D_{\mathrm{av}}^c=-1.93$, $D_{\mathrm{av}}^v=-7.67$ at $T=4$~K. Lower panels show $D_s \equiv \mathscr{D}(D_{\mathrm{an}}^c,D_{\mathrm{an}}^v,\xi_{2l}) - \gamma \sqrt{\frac32}\frac{c_l}{c_t} \mathscr{D}(D_{\mathrm{an}}^c,D_{\mathrm{an}}^v,\xi_{2t})$ with $D_{\mathrm{an}}^c=-1.67$, $D_{\mathrm{an}}^v=-1.08$. The values of $D_b$ and $D_s$ are almost independent of the temperature: at $T=300$~K they are hardly distinguishable in the plot scale.
}
\label{fig:Db}
\end{figure}

The calculated size dependences 
of $D_b \equiv \mathscr{D}(D_{\mathrm{av}}^c,D_{\mathrm{av}}^v,\xi_0)$ are shown in Fig.~\ref{fig:Db}
while the resulting size dependences of the Huang-Rhys factor 
\eqref{eq:HRb} are given in Fig.~\ref{fig:HR}. The values of the deformation potentials are taken from Ref.~\cite{rabii} (Table~X) while the band structure parameters are from Ref.~\cite{Kang1997}. The mass density of $\rho=7.6$~g/cm$^3$ is taken from Ref.~\cite{ravich}.



\section{Spheroidal mode with \texorpdfstring{$j=2$}{j=2}}\label{sec:sph1}

The displacement field for the spheroidal mode with total angular momentum $j=2$ and momentum projection $m$ can be written as (see Ref.~\cite{ufn})
\begin{widetext}
\begin{multline}\label{eq:defSM}
\hat{{\bf u}}_{2m}(\bm{r}) \equiv \hat{{\bf u}}_{2m}(r,\Theta', \phi') = \hat{A}_{2l} \, \left[ \sqrt{\frac{3}{5}} \, j_3(qr) \, {\bf Y}^3_{2m} ( \Theta', \phi' ) +
\sqrt{\frac{2}{5}} \, j_1(qr) \, {\bf Y}^1_{2m} ( \Theta', \phi' ) \right]
\\
+ \hat{A}_{2t} \, \left[ \sqrt{\frac{2}{5}} \, j_3(Qr) \, {\bf Y}^3_{2m} ( \Theta', \phi' ) -
\sqrt{\frac{3}{5}} \, j_1(Qr) \, {\bf Y}^1_{2m} ( \Theta', \phi' ) \right] \,,
\end{multline}
\end{widetext}
where ${\bf Y}^{j \pm 1}_{jm}({\bm r}/r)$ are the vector spherical functions \cite{Varshalovich}. We do not consider the possible splitting of the mode originating from elastic anisotropy (see e.g. \cite{Portales08}). Here we distinguish the axis of angular momentum quantization, $z'$, coinciding with the axis of one of the valleys, and the valley axis, $z$, which can refer to any of the valleys. The dispersion equation for the mode frequency, $\omega_0$, can be found in Ref.~\cite{ufn} and will be reproduced in Eq.~(\ref{disp_sph}). The wave numbers are related to the mode frequency via $q=\omega_0/c_l$, $Q=\omega_0/c_t$. Eq.~\eqref{eq:defSM} implies that the direction of the $z'$ axis has been chosen as the axis of the angular momentum quantization. Unlike the breathing mode, the deformation field for the spheroidal mode is not isotropic and its orientation with respect to the direction of the valley wave vector should be specified. We will align the $z$ axis along the valley wave vector and choose the two other coordinate axes in such a way that the $z'$ axis forms an angle $\beta$ with the axis $z$ in the $yz$ plane. In the coordinate frame associated with the valley, the interaction Hamiltonian takes the form 
\begin{equation}\label{eq:HdefSM}
\hat{H}^{\eta}_{def}({\bf{r}}) = 
D_{\mathrm{av}}^{\eta} \bm{\nabla\cdot \hat{{\bf u}}}_{2m} + 
D_{\mathrm{an}}^{\eta}
\left(\frac{\partial{\hat{u}^x_{2m}}}{\partial x} + \frac{\partial{\hat{u}^y_{2m}}}{\partial y} -2  \frac{\partial{\hat{u}^z_{2m}}}{\partial z} \right) 
\,.
\end{equation}
The calculation of derivatives in Eq.~\eqref{eq:HdefSM} is rather lengthy and the result may be found in Appendix~\ref{sec:Y_n_dY}. However, we have shown in Sec.~\ref{sec:matrel} that only terms with $Y_{00}$ contribute to the electron-phonon interaction, and, therefore, we can leave only these terms in \eqref{eq:HdefSM}. The result for the $\ell=0$, $\ell_z=0$ component is
\begin{equation}\label{eq:VdefSM}
    \hat{D}^{\eta, \mathrm{s}}_{00} = d_{0m}^{(2)}(\beta)\frac{2}{\sqrt5}\, D_{\mathrm{an}}^{\eta} \, 
    \left[ \hat{A}_{2l} \, q \, j_0(qr)- \sqrt{\frac32} \hat{A}_{2t} \, Q \, j_0(Qr) \right]\,,
\end{equation}
where $d_{0m}^{(2)}$ is an element of the Wigner $d$-matrix.
This should be substituted into Eq.~\eqref{eq:eph_matrel_res} to obtain the contribution from the spheroidal mode. 

For the actual calculation of the Huang-Rhys factor for the spheroidal mode, one needs to solve the dispersion equation \cite{ufn}
\begin{multline}
   \xi_{2t}^2       j_2(\xi_{2l}) j_2(\xi_{2t}) \left[ 5-\frac{\xi_{2t}^2}2 \right]
 +8\xi_{2l}\xi_{2t} j_3(\xi_{2l}) j_3(\xi_{2t})
\\
 + \xi_{2t}         j_2(\xi_{2l}) j_3(\xi_{2t}) \left[ \xi_{2t}^2 - 16 \right]
\\ 
 +2\xi_{2l}         j_3(\xi_{2l}) j_2(\xi_{2t}) \left[ \xi_{2t}^2 - 12 \right]
 = 0
\label{disp_sph}
\end{multline}
to find $\xi_{2l}\equiv qR$, $\xi_{2t}\equiv QR=(c_l/c_t)\xi_{2l}$. For PbS, the first root of this equation is $\xi_{2l}\approx 0.969$.
It is convenient to represent 
the Huang-Rhys factor as follows:
\begin{widetext}
\begin{equation}\label{eq:HR_s}
 S_2 =   \frac{\mathcal{S}_s}{R^2}
       { \left(\mathscr{D}(D_{\mathrm{an}}^c,D_{\mathrm{an}}^v,\xi_{2l}) 
       - \gamma \sqrt{\frac32}\frac{c_l}{c_t} \mathscr{D}(D_{\mathrm{an}}^c,D_{\mathrm{an}}^v,\xi_{2t}) \right)^2 } \,,
\end{equation}
where 
\begin{align}
 \mathcal{S}_s &= \frac1{10\pi\hbar\rho c_l^3}\frac{1}{\xi_{2l}A(\xi_{2l},\xi_{2t})} \,,
\\
 \gamma &= \frac{5(c_l/c_t)^2\xi_{2l}j_2(\xi_{2l}) -4j_1(\xi_{2l}) - 24j_3(\xi_{2l})}{ 2\sqrt6 (4j_3(\xi_{2t})-j_1(\xi_{2t})) } \,,
\\
  A(\xi_{2l},\xi_{2t}) &= \frac{3B(\xi_{2l})+2C(\xi_{2l})+2\gamma^2B(\xi_{2t})+3\gamma^2C(\xi_{2t})}{10}
   - 2\sqrt6 \gamma \frac{j_2(\xi_{2l})j_2(\xi_{2t})}{\xi_{2l}\xi_{2t}} \,,
\\
  B(x) &= j_3^2(x)-j_2(x)j_4(x)\,;\;\;\; C(x) = j_1^2(x)-j_0(x)j_2(x)\,,
\end{align}
\end{widetext}
and we took into account that $\sum\limits_m \left| d_{0m}^{(2)}(\beta) \right|^2=1$.

The numerical values of $\mathcal{S}_s$ at temperature 4~K and 300~K for the fundamental mode ($n=0$) and the first overtone ($n=1$) are given in Table~\ref{tbl:prefactors}. 

\begin{table}\caption{Numerical values in Eqs.~(\ref{eq:HR_b}) (\ref{eq:HR_s}).}
\label{tbl:prefactors}
\begin{tabular*}{\linewidth}{@{\extracolsep{\fill}}llrr}
 \hline
 \hline
 $T$ (K) & $n$ & $\mathcal{S}_b$ (\AA$^2/$eV$^2$) & $\mathcal{S}_s$ (\AA$^2/$eV$^2$) \\
 \hline
$4  $ & $0$ & $ 9.481$ & $14.823$ \\
$4  $ & $1$ & $18.760$ & $ 7.781$ \\
$300$ & $0$ & $11.889$ & $18.429$ \\
$300$ & $1$ & $23.519$ & $ 9.720$ \\
 \hline
 \hline
 \end{tabular*}
 \end{table}

\begin{figure*}[t]
\includegraphics[width=0.9\textwidth]{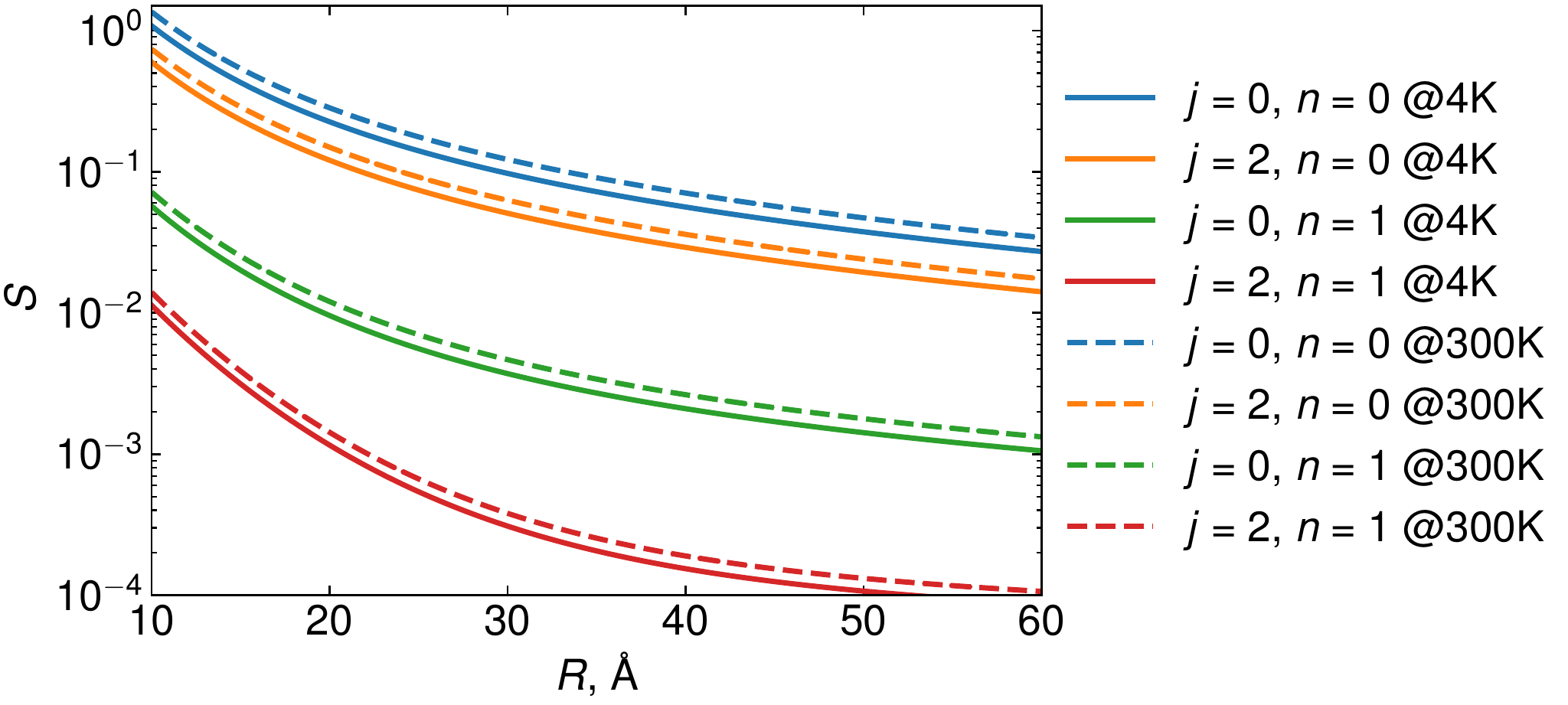}
\caption{Comparison of the single-valley Huang-Rhys factors for spheroidal vibrations of PbS QDs with $j=0$ and $j=2$.
}
\label{fig:HR}
\end{figure*}

The Huang-Rhys factors for the spheroidal and the radial breathing modes are compared in Fig.~\ref{fig:HR}. One can see that the Huang-Rhys factors for the fundamental breathing and spheroidal modes are of the same order of magnitude, though the Huang-Rhys factor is higher for the radial breathing mode.

\section{Conclusions}\label{concl}

We have analyzed exciton coupling with acoustic phonons in PbS QDs via deformation potential. We have found that interaction with the
breathing mode is due to the isotropic part of the deformation potential while interaction with the spheroidal mode with the total angular momentum $j=2$ 
is only possible due to the anisotropic part of the deformation potential. 

The values of deformation potentials for PbS used in our studies yielded comparable values of the Huang-Rhys factors for the coupling of the exciton, confined in a QD, with
the fundamental phonon modes characterized by the total angular momenta $j=0$ and $j=2$. However, these deformation potentials date back to the 1968 paper by Rabii~\cite{rabii} while 
recent state-of-the art calculations of the deformation potentials in PbTe~\cite{Murphy18} gave values very different from these of Ref.~\cite{rabii}. Thus, a thorough theoretical and experimental determination of the deformation potentials in bulk PbX is called for. In particular, one can imagine a situation when the conduction- and valence-band contributions, associated with the isotropic deformation potentials, compensate for one another in the expression for the coupling of the exciton, confined in a QD, with the breathing mode. In this case, the spheroidal mode with $j=2$, whose coupling with the confined exciton is governed by the anisotropic part of the deformation potential, would dominate low-frequency Raman spectrum, as observed by Ikezawa et al. in PbSe QDs~\cite{Ikezawa2001}.

 In this work we considered states of a confined exciton originating from a single valley. The distribution of the exciton states among the valleys can be taken into account using the extended ${\bf k \cdot p}$ theory developed in Ref.~\cite{Avdeev2020} and requires special consideration for each particular nanocrystal shape and size.

\section*{Acknowledgments}
This work was supported in part by NSF through DMR-2100248 and in part by the Russian Science Foundation Grant
No. 20-42-04405 (analytic and numerical calculations done at the Ioffe Institute).
MN also acknowledges support from the Foundation for Advancement of Theoretical Physics and Mathematics ``BASIS''.

\begin{appendix}

\section{Kang-Wise model}\label{sec:KW}

Electronic states in a spherically symmetric system can be characterized by the total angular momentum $F$ and parity. The ground state of the conduction-band electron confined in a spherical PbX QD has the total angular momentum $F_c=1/2$ and the parity $(-1)^{F_c+1/2}$~\cite{Kang1997}. The corresponding solution of Eq.~(\ref{dimmock}) can be constructed as follows. We first look for a solution of Eq.~(\ref{dimmock}) in the form
\begin{subequations}\label{sol1}
\begin{align}
\hat{u}({\bf r})&=A \, j_{F_c-1/2}(kr) \, \hat{\Omega}^{F_c-1/2}_{F_c,M_c}
\left( \frac{\bf r}{r} \right) \,,
\label{sol11}
\\
\hat{v}({\bf r})&=B \, j_{F_c+1/2}(kr) \, \hat{\Omega}^{F_c+1/2}_{F_c,M_c}
\left( \frac{\bf r}{r} \right) \,,
\label{sol12}
\end{align}
\end{subequations}
where $\hat{\Omega}^{F_c \pm 1/2}_{F_c,M_c}$ is the spherical spinor~\cite{Varshalovich} and $j_{F_c \pm 1/2}(kr)$ is the spherical Bessel function.
Using 
\begin{multline}
\left( {\bm \sigma} {\bm \nabla} \right) j_{F_c \pm 1/2} (kr) \, \hat{\Omega}^{F_c \pm 1/2}_{F_c,M_c} \\ =
\mp k \, j_{F_c \mp 1/2} (k_cr) \, \hat{\Omega}^{F_c \mp 1/2}_{F_c,M_c}
\label{identity_j}
\end{multline}
from the first of Eqs.~(\ref{dimmock}) 
we obtain
\begin{equation}
B=i \, \frac{E_g+2 \, \alpha_c \, k^2 - 2 \, E}{2 \, P \, k} \, A \equiv i \, \rho(k) \, A \,,
\end{equation}
while the second of Eqs.~(\ref{dimmock}) yields
\begin{equation}
k=\sqrt{\Pi+\Sigma} \,, 
\label{k}
\end{equation}
where
\[
\Sigma=\frac{E \, (\alpha_v-\alpha_c) - P^2 -E_g \, (\alpha_v+\alpha_c)/2}{2 \alpha_c \alpha_v} \,,
\]
\[
\Pi=\sqrt{\Sigma^2 + \frac{E^2 - (E_g/2)^2}{\alpha_c \alpha_v}} \,.
\]
Another solution of Eqs.~(\ref{dimmock})
is given by
\begin{subequations}\label{sol2}
\begin{eqnarray}
&&\hat{u}({\bf r})=C \, i^{(1)}_{F_c-1/2}(\kappa r) \, \hat{\Omega}^{F_c-1/2}_{F_c,M_c}
\left( \frac{\bf r}{r} \right) \,, \label{sol21}\\
&&\hat{v}({\bf r})=D \, i^{(1)}_{F_c+1/2}(\kappa r) \, \hat{\Omega}^{F_c+1/2}_{F_c,M_c}
\left( \frac{\bf r}{r} \right) \,. \label{sol22}
\end{eqnarray}
\end{subequations}
Using 
\begin{multline}
\left( {\bm \sigma} {\bm \nabla} \right) i^{(1)}_{F_c \pm 1/2} (\kappa r) \, \hat{\Omega}^{F_c \pm 1/2}_{F_c,M_c}\\=
- \kappa \, i^{(1)}_{F_c \mp 1/2} (\kappa r) \, \hat{\Omega}^{F_c \mp 1/2}_{F_c,M_c}
\label{identity_i}
\end{multline}
from the first of Eqs.~(\ref{dimmock}) 
we obtain
\begin{equation}
D=i \, \frac{E_g-2 \, \alpha_c \, \kappa^2 - 2 \, E}{2 \, P \, \kappa} \, C \equiv i \, \mu(\kappa) \, C\,,
\end{equation}
while the second of Eqs.~(\ref{dimmock}) yields
\begin{equation}
\kappa=\sqrt{\Pi-\Sigma} \,.
\label{kappa} 
\end{equation}
From the condition that a linear combination of the solutions~(\ref{sol11}),~(\ref{sol12}) on one hand and the solutions~(\ref{sol21}),~(\ref{sol22})
on the other hand vanishes at $r=R$, where $R$ is the QD radius,
we obtain the dispersion equation for $k \equiv k_c$, $\kappa \equiv \kappa_c$~\cite{Kang1997}
\begin{multline}
i^{(1)}_{F+1/2}(\kappa_c R) \,
j_{F-1/2}(k_c R) \,
\mu(\kappa_c) 
\\
-i^{(1)}_{F-1/2}(\kappa_c R) \,
j_{F+1/2}(k_c R) \,
\rho(k_c)=0 \,,
\end{multline}
which yields the energy of the confined conduction-band electron state ($E>0$).
Thus, for the ground state of the conduction-band electron confined in a spherical PbX QD
we have a bispinor wave function
\begin{subequations}
\begin{align}
\hat{u}^c_{1/2,M_c}({\bf r})&=R^{-3/2} \, z_0^c \left( \frac{r}{R} \right) \, \hat{\Omega}^0_{1/2,M_c}
\left( \frac{\bf r}{r} \right) \,,
\label{uce}
\\
\hat{v}^c_{1/2,M_c}({\bf r})&=i \, R^{-3/2} \, z_1^c \left( \frac{r}{R} \right) \, \hat{\Omega}^1_{1/2,M_c}
\left( \frac{\bf r}{r} \right) \,.
\label{vce}
\end{align}
\end{subequations}
The radial wave functions are~\cite{Kang1997}
\begin{widetext}
\begin{align}
z^c_{F_c-1/2}(x)&=A_c \left[ j_{F_c-1/2}(k_c R \, x) -\frac{j_{F_c-1/2}(k_c R)}{i^{(1)}_{F_c-1/2}(\kappa_c R)}
\, i^{(1)}_{F_c-1/2}(\kappa_c R \, x) \right] \,,
\label{z0c}
\\
z^c_{F_c+1/2}(x)&=A_c \left[ \rho(k_c) \, j_{F_c+1/2}(k_c R \, x) - \mu(\kappa_c) \, \frac{j_{F_c-1/2}(k_c R)}{i^{(1)}_{F_c-1/2}(\kappa_c R)}
\, i^{(1)}_{F_c+1/2}(\kappa_c R \, x) \right] \,,
\label{z1c}
\end{align}
\end{widetext}
where $A_c$ is a normalization constant determined by the condition
\[
\int\limits_0^1 dx x^2 \left[ z_{F_c-1/2}^{c \, 2} (x)
+  z_{F_c+1/2}^{c \, 2} (x) \right]=1 \,.
\]
Apart from the explicit dependence on $x$ these functions weakly depend on $R$. In this work we will only need them for $F_c=1/2$.

The ground state of the valence-band hole confined in a spherical PbX QD has the total angular momentum $F_h=1/2$ and the parity $(-1)^{F_h-1/2}$~\cite{Kang1997}. We will again construct two solutions of Eq.~(\ref{dimmock}) in a free space and impose a boundary condition on their linear combination. This time for the first solution of Eqs.~(\ref{dimmock}) we use the substitution
\begin{subequations}
\begin{align}
\hat{u}({\bf r})&=A \, j_{F_h+1/2}(kr) \, \hat{\Omega}^{F_h+1/2}_{F_h,M_h}
\left( \frac{\bf r}{r} \right) \,,
\\
\hat{v}({\bf r})&=B \, j_{F_h-1/2}(kr) \, \hat{\Omega}^{F_h-1/2}_{F_h,M_h}
\left( \frac{\bf r}{r} \right) \,,
\end{align}
\end{subequations}
and get
\[
B=-i \, \rho(k) \, A \,.
\]
For the second solution of Eqs.~(\ref{dimmock}) we try
\begin{subequations}
\begin{align}
\hat{u}({\bf r})&=C \, i^{(1)}_{F_h+1/2}(\kappa r) \, \hat{\Omega}^{F_h+1/2}_{F_h,M_h}
\left( \frac{\bf r}{r} \right) \,,
\\
\hat{v}({\bf r})&=D \, i^{(1)}_{F_h-1/2}(\kappa r) \, \hat{\Omega}^{F_h-1/2}_{F_h,M_h}
\left( \frac{\bf r}{r} \right) \,,
\end{align}
\end{subequations}
and we again obtain
\[
D=i \, \mu(\kappa) \, C \,.
\]
From the condition that a linear combination of these two solutions vanishes at $r=R$
we obtain the dispersion equation for $k=k_v$ and $\kappa=\kappa_v$~\cite{Kang1997} 
\begin{multline}
i^{(1)}_{F_h-1/2}(\kappa_v R) \,
j_{F_h+1/2}(k_v R) \,
\mu(\kappa_v) 
\\
+i^{(1)}_{F_h+1/2}(\kappa_v R) \,
j_{F-1/2}(k_v R) \,
\rho(k_v)=0 \,,
\end{multline}
which yields the energy of the confined valence-band hole state ($E<0$).
The resulting bispinor wave function for the ground state of the valence-band hole confined in a spherical PbX QD
takes the form 
\begin{subequations}
\begin{align}
\hat{u}^v_{1/2,M_h}({\bf r})&=R^{-3/2} \, z_1^v \left( \frac{r}{R} \right) \, \hat{\Omega}^1_{1/2,M_h}
\left( \frac{\bf r}{r} \right) \,,
\label{uvh}
\\
\hat{v}^v_{1/2,M_h}({\bf r})&=-i \,  R^{-3/2} \, z_0^v \left( \frac{r}{R} \right) \, \hat{\Omega}^0_{1/2,M_h}
\left( \frac{\bf r}{r} \right) \,.
\label{vvh}
\end{align}
\end{subequations}
The radial wave functions are~\cite{Kang1997}
\begin{widetext}
\begin{subequations}
\begin{align}
z^v_{F_h+1/2}(x)&=B_v \, \left[j_{F_h+1/2}(k_v R \, x) -\frac{j_{F_h+1/2}(k_v R)}{i^{(1)}_{F_h+1/2}(\kappa_v R)}
\, i^{(1)}_{F_h+1/2}(\kappa_v R \, x) \right] \,,
\\
z^v_{F_h-1/2}(x)&=B_v \, \left[ \rho(k_v) \, j_{F_h-1/2}(k_v R \, x) + \mu(\kappa_v) \, \frac{j_{F_h+1/2}(k_v R)}{i^{(1)}_{F_h+1/2}(\kappa_v R)}
\, i^{(1)}_{F_h-1/2}(\kappa_v R \, x) \right] \,,
\end{align}
\end{subequations}
\end{widetext}
where $B_v$ is a normalization constant determined by the condition
\[
\int\limits_0^1 dx x^2 \left[ z_{F_h-1/2}^{v \, 2} (x)
+  z_{F_h+1/2}^{v \, 2} (x) \right]=1 \,.
\]

\section{Normalization of the Radial breathing mode}\label{sec:ph_br}
The procedure of second quantization for the radial breathing mode can be done in a close analytical form.
The density of elastic energy is given by
\begin{multline}
{\cal E}_{def}=\rho \, c_t^2 \sum_{\alpha} (\bm{\nabla} u_{\alpha})^2
  -\frac{\rho \, c_t^2}{2} \left[\bm{\nabla} \times {\bf u}\right]^2 \\ +
   \rho \, \left(\frac{c_l^2}{2}-c_t^2 \right) \left(\bm{\nabla}\cdot {\bf u}\right)^2 \,.
\end{multline}
For the radial breathing mode $\left[\bm{\nabla} \times {\bf u}\right]=0$.
Displacement field for this mode (either fundamental or overtone, depending on the value of the discrete wave number $q$, cf. Eq.~(\ref{eq:dispbreath})) can be written as
\begin{equation}\label{eq:u_def_bm}
\hat{\bf u}(\bm{r})=\frac{Q_q \, j_1(qr) \, Y_{00} \, {\bf e}_r}{\sqrt{\int\limits_0^R dr \, r^2 \, j_1^2(qr)}} \,,
\end{equation}
where $Q_q$ stays for a generalized coordinate. Then the elastic density is proportional to
\[
{\cal E}_{def} \propto \frac{c_l^2}{2} \, j_0^2(qr) +\frac{2}{3} \, c_t^2 \, \left( j_2^2(qr)-j_0^2(qr) \right) \,.
\]

In order to compute the total elastic energy we need several integrals:
\begin{equation}
\int\limits_0^R dr \, r^2 \, j_1^2(qr)=\frac{1}{q^3} \, \left(\frac{qR}{2}-\frac{\sin^2{qR}}{qR}+\frac{\sin{qR} \, \cos{qR}}{2} \right) \,,
\label{eq:normint0}
\end{equation}
\[
\int\limits_0^R dr \, r^2 \, j_0^2(qr)=\frac{1}{q^3} \, \left(\frac{qR}{2}-\frac{\sin{qR} \, \cos{qR}}{2} \right) \,,
\]
\[
\int\limits_0^R dr \, r^2 \, \left( j_2^2(qr)-j_0^2(qr) \right)=- \frac{3 \,(qR)}{q^3} \, j_1^2(qR) \,. 
\]
The dispersion equation for the breathing mode \cite{ufn} can be written as
\begin{equation}
c_t^2 \, j_1(qR)=\frac{c_l^2}{4} \, qR \, j_0(qR) \,.
\label{eq:dispbreath}
\end{equation}
This allows one to rewrite the last integral as
\begin{multline*}
c_t^2 \, \int\limits_0^R dr \, r^2 \, \left( j_2^2(qr)-j_0^2(qr) \right)\\=
- c_l^2 \, \frac{3 \,(qR)^2}{4 \,q^3} \, j_0(qR) \, j_1(qR) 
\\
=
- c_l^2 \, \frac{3}{4 \,q^3} \, \left(\frac{\sin^2{qR}}{qR}-\sin{qR} \, \cos{qR} \right)\,. 
\end{multline*}
Then for the total elastic energy one obtains
\[
E_{def}=\frac{c_l^2}{2} \, \rho \, q^2 \, Q_q^2 = \frac{\rho \, \omega_0^2 \, Q_q^2}{2}
\]
while for the total kinetic energy
\[
E_{kin}=\frac{\rho}{2} \, \dot{Q}_q^2 \equiv \frac{P_q^2}{2 \rho} \,,
\]
where $P_q$ is the generalized momentum. This allows one to write
\begin{align}\label{eq:Q_bm}
Q_q&=\sqrt{\frac{\hbar}{2 \, \rho \, \omega_0}} \, \left( \hat{a}^{\dag}+\hat{a} \right) \,,
\\\label{eq:P_bm}
P_q&=i \, \sqrt{\frac{\hbar \, \rho \, \omega_0}{2}} \, \left( \hat{a}^{\dag}-\hat{a} \right) \,.
\end{align}
Equations~\eqref{eq:u_def_bm}, \eqref{eq:normint0}, \eqref{eq:Q_bm} give the normalization coefficient for the breathing mode.

\section{Normalization of the Spheroidal phonons with \texorpdfstring{$j=2$}{j=2}}\label{sec:ph_sph}

Here we give the results only.
The displacement field for the spheroidal phonon mode with $j=2$ is given by
\begin{widetext}
\begin{multline}
\hat{\bf u}_2(\bm{r})= \frac{1}{\sqrt{A(qR,QR)} \, R^{3/2}} \, \sqrt{\frac{\hbar}{2 \, \rho \, \omega_0}} \, \sum_m \left( \hat{a}_{2m}+(-1)^m \, \hat{a}^{\dag}_{2-m}
\right)
\\
\times
\Bigg\{
\left[ \sqrt{\frac{3}{5}} \, j_3(qr) \, {\bf Y}^3_{2m} \left(\frac{{\bm r}}{r} \right) +
\sqrt{\frac{2}{5}} \, j_1(qr) \, {\bf Y}^1_{2m} \left(\frac{{\bm r}}{r} \right) \right]
\\
+ \frac{b}{a} \, \left[ \sqrt{\frac{2}{5}} \, j_3(Qr) \, {\bf Y}^3_{2m} \left(\frac{{\bm r}}{r} \right) -
\sqrt{\frac{3}{5}} \, j_1(Qr) \, {\bf Y}^1_{2m} \left(\frac{{\bm r}}{r} \right) \right] 
\Bigg\}
\,,
\end{multline}
where
\begin{align*}
\frac{b}{a}&=\frac{5 \, c_l^2 \,qR \, j_2(qR) - 4 \, c_t^2 \, j_1(qR)-24 \, c_t^2 \, j_3(qR)}{2 \, c_t^2 \, \sqrt{6} \, \left(4 \, j_3(QR)-j_1(QR) \right)}
\,,
\\
A(qR,QR)&=3 \, B(qR)+2 \, C(qR) + \frac{b^2}{a^2} [2 \, B(QR) +3 \, C(QR)]-2 \, \sqrt{6} \, \frac{b}{a} \, \frac{j_2(qR) \, j_2(QR)}{qR \, QR} \,,
\\
B(x)&=\frac{j_3^2(x)-j_2(x) \, j_4(x)}{10} \,,
\\
C(x)&=\frac{j_1^2(x)-j_0(x) \, j_2(x)}{10} \,.
\end{align*}
\end{widetext}

\section{Spherical vectors and their derivatives}\label{sec:Y_n_dY}

For the calculations of deformation potentials it is convenient to use the equations for derivatives of spherical vectors. This appendix summarizes how these equations may be obtained. We use the following relation to compute differential operators of interest:
\begin{subequations}\label{eq:derivatives_explicit}
\begin{align}
    \bm{\nabla} \bm{\cdot} {\bf u} = & \nabla_+{u}^{+1}
                        + \nabla_0{u}^{0}
                        + \nabla_-{u}^{-1}
\\
    \bm{\nabla}_{z^2} \bm{\cdot} {\bf u} \equiv & \frac{\partial u_x}{\partial x} + \frac{\partial u_y}{\partial y} 
    -2  \frac{\partial u_z}{\partial z} \nonumber
    \\& = \nabla_+{u}^{+1}
     -2\nabla_0{u}^{0}
     + \nabla_-{u}^{-1}
\end{align}
\end{subequations}
where ${\bf u}$ is a vector field, and $u^{\mu}$ ($\mu=\pm1,0$) are the $\mu$-th circular contravariant components of this vector field.

We use Eq.~(45), Sec.~7.3.4 of \cite{Varshalovich} which relates spherical vectors in a rotated system $S'$ with spherical vectors in the original coordinate system. The deformation field of the spheroidal mode is a linear combination of $j_1(kr){\bf Y}_{2m}^1(\Theta',\phi')$ and $j_3(kr){\bf Y}_{2m}^3(\Theta',\phi')$ \textsl{in the phonon coordinate system} $S'$ which we obtain by the rotation of the valley coordinate system $S$ in the $xz$ plane by the angle $\beta$. We do not need to consider other Euler angles as the valley (phonon mode) are isotropic in the $xy$ ($x'y'$) plane.\footnote{The rotation $\gamma\neq0$ changes phase of spinors, which may contribute to the phase difference of the matrix elements between different valleys. However, this does not result in any observable effect unless valley mixing or intervalley transitions are considered.} As a result, the deformation amplitude contains 
\begin{multline}
    {\bf Y}_{2m}^{L}(\Theta',\phi') \\ = \sum_{M,\mu=\pm1,0} d_{Mm}^{2}(\beta) C_{LM-\mu1\mu}^{2M} 
    Y_{LM-\mu}(\Theta,\phi) {\bf e}_{\mu}\,,
\end{multline}
where $d_{Mm}^2(\beta)$ are the components of Wigner $d$-matrix.
Quite lengthy, but straightforward calculations lead to the following result:

\begin{widetext}
\begin{subequations}\label{eq:dY20_rot}
\begin{multline}\label{eq:dY201_rot}
  \bm{\nabla \cdot} j_1(kr) {\bf Y}_{2m}^1(\Theta',\phi') = - \sqrt{\frac25} kj_2(kr) 
    \bigg[ d_{0m}^2(\beta) Y_{20}(\Theta,\phi)
    -4 d^2_{1m}(\beta) Y_{2 1}(\Theta,\phi)
    \\
    +4 d^2_{1m}(\beta) Y_{2-1}(\Theta,\phi)
    -2 d^2_{2m}(\beta) Y_{2 2}(\Theta,\phi)
    -2 d^2_{2m}(\beta) Y_{2-2}(\Theta,\phi)
    \bigg]
\end{multline}
\begin{multline}\label{eq:dzY201_rot}
  \bm{\nabla}_{z^2}\bm{\cdot}  j_1(kr) {\bf Y}_{2m}^1(\Theta',\phi') = 
    -\sqrt2 kj_0(kr) d_{0m}^2(\beta) Y_{00}(\Theta,\phi) 
   \\
+ \sqrt{\frac25} kj_2(kr) 
    \bigg[ d_{0m}^2(\beta) Y_{20}(\Theta,\phi)
    -d_{1m}^2(\beta) Y_{2 1}(\Theta,\phi)    
    +d_{1m}^2(\beta) Y_{2-1}(\Theta,\phi)    
    \\
    -2 d_{2m}^2(\beta) Y_{2 2}(\Theta,\phi)
    -2 d_{2m}^2(\beta) Y_{2-2}(\Theta,\phi)
    \bigg]
\end{multline}
\begin{multline}\label{eq:dY203_rot}
  \bm{\nabla \cdot} j_3(kr) {\bf Y}_{2m}^3(\Theta',\phi') = 
    -\sqrt{\frac35} kj_2(kr) 
    \bigg[ d_{0m}^2(\beta) Y_{20}(\Theta,\phi)
    -2 d_{1m}^2(\beta) Y_{2 1}(\Theta,\phi)
    \\
    +2 d_{1m}^2(\beta) Y_{2-1}(\Theta,\phi)
    +2 d_{2m}^2(\beta) Y_{2 2}(\Theta,\phi)
    +2 d_{2m}^2(\beta) Y_{2-2}(\Theta,\phi)
    \bigg]
\end{multline}
\begin{multline}\label{eq:dzY203_rot}
  \bm{\nabla}_{z^2}\bm{\cdot}  j_3(kr) {\bf Y}_{2m}^3(\Theta',\phi') = 
    \frac{2\sqrt3}{7\sqrt5} kj_2(kr) 
    \Bigg[ d_{0m}^2(\beta) Y_{20}(\Theta,\phi)
    -d_{1m}^2(\beta) Y_{2 1}(\Theta,\phi)
    \\
    +d_{1m}^2(\beta) Y_{2-1}(\Theta,\phi) 
    -2d_{2m}^2(\beta) Y_{2 2}(\Theta,\phi)
    -2d_{2m}^2(\beta) Y_{2-2}(\Theta,\phi)
    \Bigg]
    \\
    -\frac{4\sqrt3}{7} kj_4(kr) 
    \Bigg[ d_{0m}^2(\beta) Y_{40}(\Theta,\phi)
    -\sqrt{\frac{10}3} d_{1m}^2(\beta) Y_{4 1}(\Theta,\phi)
    \\
    +\sqrt{\frac{10}3} d_{1m}^2(\beta) Y_{4-1}(\Theta,\phi)
    +\sqrt{\frac53} d_{2m}^2(\beta) Y_{4 2}(\Theta,\phi)
    +\sqrt{\frac53} d_{2m}^2(\beta) Y_{4-2}(\Theta,\phi)
    \Bigg]\,.
\end{multline}
\end{subequations}
\end{widetext}


\end{appendix}


\bibliography{PbS}

\end{document}